# Cyclo-Synchrotron Emission and Comptonization in Optically Thin Hot Accretion Flows

Rohan Mahadevan

Harvard-Smithsonian Center for Astrophysics, 60 Garden St., Cambridge, MA 02138, USA

September 26, 1995

**Abstract.** Soft cyclo-synchrotron photons produced by thermal electrons moving in a magnetic field can be subsequently Comptonized to produce hard X-ray and $\gamma$-ray emission. An accurate determination of the amount of soft cyclo-synchrotron photons produced from such plasmas is clearly necessary. We present the complete solution to the cyclo-synchrotron problem. We calculate the emission produced by an isotropic distribution of thermal electrons moving in a magnetic field, for temperatures ranging from $5 \times 10^8$K to $3.2 \times 10^{10}$K. The applicability of the standard synchrotron formula is evaluated, and we provide convenient fitting functions for the emission spectra within this range of temperatures. We discuss how the Comptonized spectra, for various temperatures and optical depths, are affected by the new results.

## 1. Introduction

Radiation by quasirelativistic and fully relativistic electrons in magnetic fields is very common in astrophysics. Examples of sources include jets and lobes in radio galaxies (e.g. Carilli et al. 1991), hot accretion flows onto neutron stars and black holes (e.g. Narayan, Yi & Mahadevan 1995), and relativistic fireballs in gamma-ray bursts (e.g. Paczyński & Rhoads 1993; Mészáros, Laguna & Rees 1993). These are a few examples of where we encounter thermal or non-thermal electrons radiating cyclo-synchrotron radiation in a magnetic field which is probably of near-equipartition strength. Under certain circumstances (e.g. Takahara & Tsuruta 1982; Narayan, Yi & Mahadevan 1995) this radiation can then be multiply Compton scattered to give rise to detailed high energy spectra from these objects. The ability to accurately calculate cyclo-synchrotron emission is clearly important to the study and understanding of these systems.

Surprisingly, only some limiting results are available in this field. Previous results (Schwinger 1949; Rosner 1958; Oster 1960, 1961; Bekefi 1966; Pacholczyk 1970) have considered extreme cyclotron or synchrotron emission where analytical formulae have been used. These calculations do not treat the transition from cyclotron to synchrotron radiation as the velocities of the electrons become quasirelativistic, and can drastically overestimate the total amount of emission produced unless we are able to solve the problem exactly. Previous work in the mildly relativistic regime (Petrosian 1981; Takahara & Tsuruta 1982; Melia 1994) include approximations which are not assumed here (see Mahadevan, Narayan & Yi 1996). As far as we know a complete treatment of the problem has not be presented so far.

## 2. Theory

We give an outline of the complete solution (see Mahadevan, Narayan & Yi 1996 for details). The power (energy/time/steradian/frequency) emitted by an electron moving with a velocity parameter between $\boldsymbol{\beta}$ and $\boldsymbol{\beta} + d\boldsymbol{\beta}$, in a frequency range $d\omega$, and at an observer angle $\theta$, is given by (Schott 1912; Rosner 1958; Bekefi 1966),

$$\eta_\omega(\boldsymbol{\beta}, \theta)\, d\omega = \frac{e^2 \omega^2}{2\pi c}$$
$$\times \left[\sum_{m=1}^{\infty} \left(\frac{\cos\theta - \beta_\parallel}{\sin\theta}\right)^2 J_m^2(x) + \beta_\perp^2 J_m'^2(x)\right] \delta(y)\, d\omega, \quad (1)$$

where

$$x = \frac{\omega}{\omega_o} \beta_\perp \sin\theta, \quad (2)$$

$$y = m\omega_o - \omega(1 - \beta_\parallel \cos\theta), \quad (3)$$

$$\omega_o = \frac{\omega_b}{\gamma} = \frac{eB}{\gamma m_e c}, \quad (4)$$

$\delta(y)$ is the Dirac delta function, $J_m(x)$ is the Bessel function of order $m$, $J'_m(x)$ is the derivative of the Bessel function, $\beta_\parallel = \beta \cos\theta_p$, and $\beta_\perp = \sin\theta_p$ is the velocity parameter parallel and perpendicular to the magnetic field, $B$, respectively. $\gamma = \sqrt{1-\beta^2}$ is the Lorentz factor.

Each integer $m$ in the summation in Eq.(1) corresponds to a harmonic. To obtain the emission in the cyclotron limit (low $\beta$), we expand the Bessel functions in Eq.(1) to lowest order and obtain (Schwinger 1949; Rosner 1956; Bekefi 1966)

$$\eta_m^T = \frac{2e^2\omega_b^2}{c} \frac{(m+1)(m^{2m+1})}{(2m+1)!} \beta^{2m}. \tag{5}$$

Fig. 1a shows an example with $\beta = 0.3$, which corresponds to this limit. The widths of each harmonic in Fig. 1a is due to artificially broadening the delta function in Eq.(1) to facilitate the numerics.

In the opposite limit corresponding to the extreme synchrotron case ($\gamma \gg 1$), we are interested in large $m$'s and Eq.(1) reduces to (Schwinger 1949; Bekefi 1966; Rybicki & Lightman 1979)

$$\frac{dE}{d\omega} = \frac{\sqrt{3}e^3 B \sin\theta_p}{2\pi m_e c^2} F(x), \tag{6}$$

where $x = \omega/\omega_c$, with $\omega_c = (3/2)\gamma^2 \omega_b \sin\theta_p$, $F(x)$ is given by

$$F(x) \equiv x \int_x^\infty K_{\frac{5}{3}}(\xi)\,d\xi, \tag{7}$$

and $K_{\frac{5}{3}}(\xi)$ is the modified Bessel function. Fig. 1d shows an example with $\gamma = 20$, which corresponds to this limit.

These results are valid for electrons of a given velocity. For a thermal distribution of electrons, we need to integrate the emission over a Maxwellian. This integral can be done in the synchrotron limit, for a fixed particle direction, $\sin\theta_p$, to give (Pacholczyk 1970),

$$\begin{aligned}\varepsilon_\omega d\omega &= 4\pi \times \frac{e^2}{(2\pi)^2\sqrt{3}c} \frac{n_e \omega}{K_2(1/\theta_e)} \\ &\quad \times I\left(\frac{x_M}{\sin\theta_p}\right) d\omega \quad \text{ergs cm}^{-3}\text{ s}^{-1}\text{ Hz}^{-1},\end{aligned} \tag{8}$$

where $x_M = 2\omega/3\omega_o\theta_e^2$, $\theta_e = kT/m_e c^2$, $K_2(1/\theta_e)$ is the modified Bessel function, and $I(X)$ is given by

$$I(X) \equiv \frac{1}{X} \int_0^\infty z^2 \exp(-z) F(X/z^2) dz. \tag{9}$$

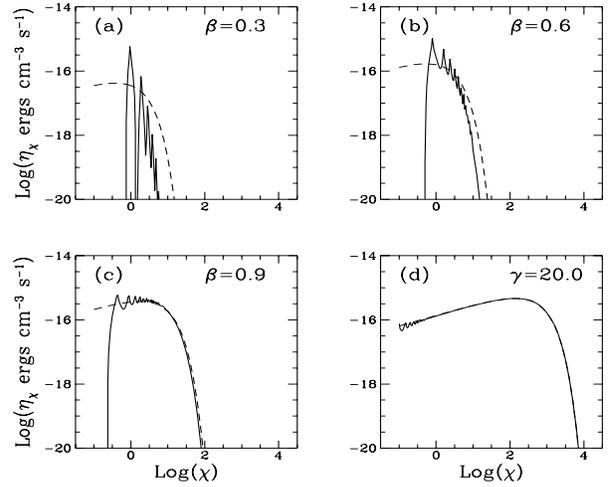

**Fig. 1.** Emission produced by a single electron moving with (a) $\beta = 0.3$, (b) $\beta = 0.6$, (c) $\beta = 0.9$, (d) $\gamma = 20.0$. Dashed curves represent the synchrotron approximation Eqn. (6). The x-axis is the log of the dimensionless frequency $\chi = \omega/\omega_b$, and the y-axis is $\eta_\chi = \eta_\omega \omega_b$

To obtain the exact emission spectrum for an isotropic thermal distribution of electrons, $n(\beta)$, for any temperature, we evaluate $L_\omega \equiv dE/d\omega$ given by

$$\begin{aligned}L_\omega \equiv \frac{dE}{d\omega} &= \frac{2}{4\pi} \int_0^1 d\beta\, n(\beta) \int_0^{2\pi} d\phi_p \\ &\quad \times \int_0^1 d(\cos\theta_p) \int_0^{2\pi} d\phi \int_{-1}^1 d(\cos\theta)\, \eta_\omega(\beta,\theta).\end{aligned} \tag{10}$$

For a fixed $\beta$ and $\omega$, we numerically evaluate Eq.(10), as well as the sum over all the harmonics in Eq.(1). We repeat this calculation for various values of $\beta$ and $\omega$ and tabulate the results. The results can then be convolved with any isotropic velocity distribution to obtain the spectrum $L_\omega$. In this paper we restrict ourselves to a relativistic Maxwellian $n(\beta)$ and present results for this particular case.

## 3. Results

### 3.1. Harmonic Nature of Synchrotron Emission

Fig. 1 shows the total emission produced by a single electron moving with different velocities. We plot $\eta_\chi = dE/d\chi$ against $\chi$, where $\chi \equiv \omega/\omega_b$ is the dimensionless frequency. In these units the vertical axis scales with the magnetic field, $B$, and the horizontal axis remains unchanged. The dashed curves are the synchrotron approximation Eq.(6). Even at high values of $\gamma$, the synchrotron approximation overestimates the emission at low frequencies $\chi$. This is due to discrete harmonic emission.

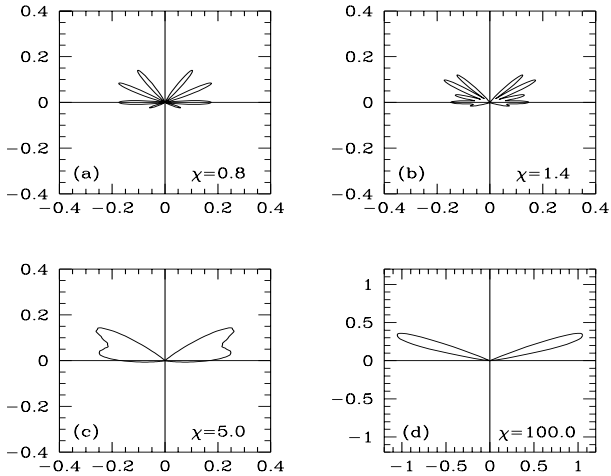

**Fig. 2.** Polar plots for a particle with $\gamma = 20$, moving at an angle of $\cos\theta_p = 0.3$ to the magnetic field. The magnetic field is oriented along the positive y-axis. Different plots correspond to observing the emission at different frequencies (a) $\chi = 0.8$, (b) $\chi = 1.4$, (c) $\chi = 5.0$, (d) $\chi = 100.0$.

Fig. 2 shows polar plots of emission, at different $\chi$'s, for a particle moving with a Lorentz factor $\gamma = 20.0$ (cf. Fig. 1d). At high values of $\chi$ (Fig. 2d) we see the relativistic beaming effect which is what we expect in the synchrotron limit. This is where the synchrotron approximation and the numerical calculation agree. Discrete harmonic emission is dominant at low frequencies where we obtain "bumps" in the spectrum, and the agreement with the synchrotron approximation is poor (see Mahadevan, Narayan & Yi 1996).

### 3.2. Integration Over a Thermal Distribution of Electrons

We have computed emission spectra for isotropic thermal distributions of electrons with temperatures in the range, $5 \times 10^8 \text{K} < T < 3.2 \times 10^{10} \text{K}$. The solid lines in Fig. 3 represent the numerical calculations, and the dashed lines represent the synchrotron approximation, Eq.(8), after averaging over particle directions. It is evident that the synchrotron approximation for the emission overestimates the true emission at low temperatures, but is a good approximation to the true emission for $T \gtrsim 10^{10} \text{K}$, as expected. We have obtained a set of fitting functions corresponding to each of the temperatures for which we have calculated the spectrum. The functions are of the form

$$M(x_M) = \frac{4.0505\alpha}{x_M^{1/6}}\left(1 + \frac{0.40\beta}{x_M^{1/4}} + \frac{0.5316\gamma}{x_M^{1/2}}\right)$$
$$\times \exp\left(-1.8896\, x_M^{1/3}\right), \qquad (11)$$

with $M(x_M)$ replacing $I(x_M)$ in Eq.(8). With this substitution Eq.(8) represents the total emission from an isotropic thermal distribution of electrons. In Eq.(11),

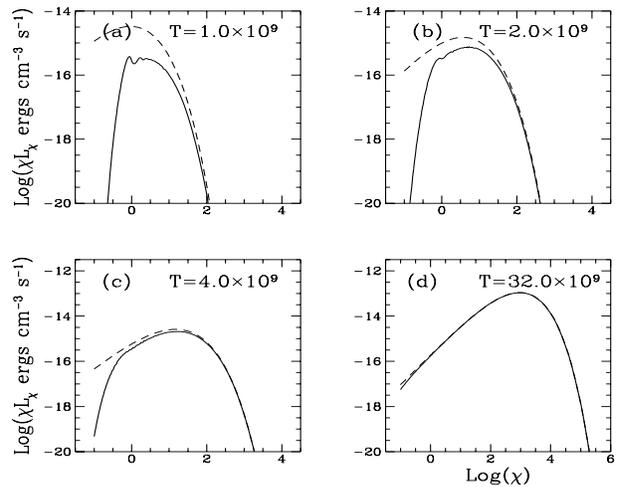

**Fig. 3.** Plots comparing the synchrotron approximation (dashed lines), Eqn. (8) (after averaging over particle directions), with the exact numerical results (solid lines), for the emission from a thermal distribution of isotropic electrons at different temperatures. (a) $T = 1.0 \times 10^9$, (b) $T = 2.0 \times 10^9$, (c) $T = 4.0 \times 10^9$, (d) $T = 32.0 \times 10^9$.

$\alpha$, $\beta$, $\gamma$, are adjustable parameters that are optimized so as to minimize the square of the deviation of $M(x_M)$ from the numerically calculated results. Table 1 shows the optimized parameters for various temperatures. Fig. 4 compares the fitted functions (dashed lines), with the numerically calculated results (solid lines). The maximum error in the fits range from 10% at $10^9 K$, to 0.1% at $3.2 \times 10^{10} K$.

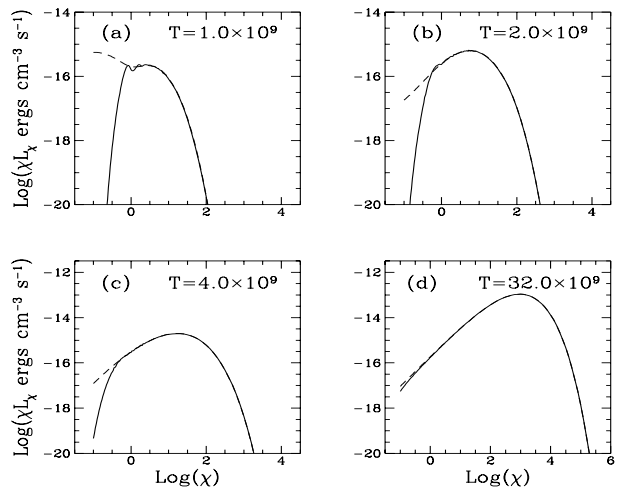

**Fig. 4.** Plots comparing our fitting functions (dashed lines) with the exact numerical results (solid lines), for the emission from a thermal distribution of isotropic electrons at different temperatures. (a) $T = 1.0 \times 10^9$, (b) $T = 2.0 \times 10^9$, (c) $T = 4.0 \times 10^9$, (d) $T = 32.0 \times 10^9$.

## 4. Effects on the Comptonized Spectrum

For certain temperatures and frequency ranges, Fig. 3 shows how the synchrotron approximation overestimates the exact calculation. For plasmas where the optical depth $\tau \lesssim 1$ for all frequencies, this overestimate can drastically effect the amount of Compton scattered photons in the X-rays and $\gamma$-rays. In plasmas where the emission is self absorbed for $\log(\chi) \lesssim 2$, we are only interested in the cyclo-synchrotron emission for $\log(\chi) > 2$ (Petrosian 1981; Takahara & Tsuruta 1982; Narayan, Yi & Mahadevan 1995). Fig. 3 shows that the synchrotron approximation is a fairly good approximation to the exact calculation, in this regime, and the results presented do not effect the amount of Compton scattered photons.

As an example we calculate the emission produced by accretion onto a $7.0 \times 10^5 M_\odot$ black hole with $\dot{M} = 10^{-4} \dot{M}_{Edd}$. We assume an isothermal sphere of radius 100 Schwarzschild, with a uniform number density, magnetic field, and temperature. We calculate cyclo-synchrotron emission, Comptonization of the soft cyclo-synchrotron photons by the thermal electrons, bremsstrahlung emission, and Comptonized bremsstrahlung emission. Fig. 5a shows the emission spectrum for a temperature of $10^9$K. The first peak corresponds to synchrotron emission, and the second to bremsstrahlung emission. Comptonization is not important at this temperature. In Figs. 5b, c, d, we see the Compton scattered emission, as the rising peak in the middle, as well as the Comptonized bremsstrahlung emission as a rise in the falling bremsstrahlung emission at the high frequency end of the spectrum.

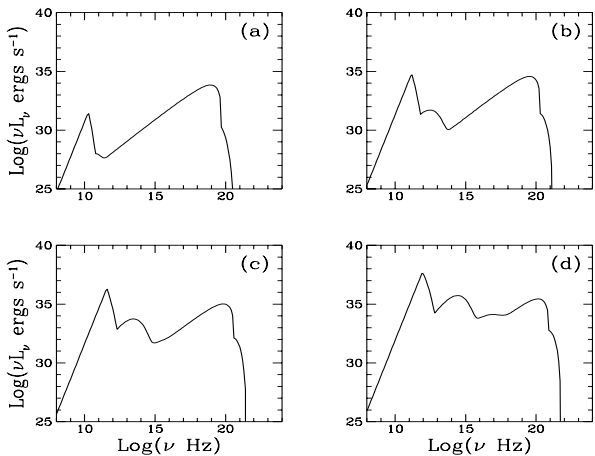

**Fig. 5.** Emission produced by accretion onto a $7.0 \times 10^5 M_\odot$ black hole with $\dot{M} = 10^{-4} \dot{M}_{Edd}$. (a) $T = 1.0 \times 10^9$, (b) $T = 4.0 \times 10^9$, (c) $T = 8.0 \times 10^9$, (d) $T = 16.0 \times 10^9$.

## 5. Conclusion

We have obtained the complete solution to cyclo-synchrotron emission produced by an isotropic distribution of thermal electrons moving in a magnetic field. For $T \lesssim 3 \times 10^{10}$K, standard synchrotron approximations overestimate the true emission. This discrepancy is important when incorporating Comptonization for optically thin plasmas, since in some models (e.g. Narayan, Yi & Mahadevan 1995) it is the soft cyclo-synchrotron photons that are Comptonized by the thermal electrons. However, when self absorption is important, the Comptonized spectra are unchanged by the new results since the emission, at frequencies where the synchrotron approximation grossly overestimates the true calculation, is self absorbed.

*Acknowledgements.* This work was supported in part by NSF grant AST 9423209. The author thanks Ramesh Narayan and Insu Yi for useful discussions and comments.

**Table 1.** Optimal values of the parameters for different temperatures.

| $T$ (K) | $I'(x)$ | | |
|---|---|---|---|
| | $\alpha$ | $\beta$ | $\gamma$ |
| $5 \times 10^8$ | 0.0431 | 10.44 | 16.61 |
| $1 \times 10^9$ | 1.121 | -10.65 | 9.169 |
| $2 \times 10^9$ | 1.180 | -4.008 | 1.559 |
| $4 \times 10^9$ | 1.045 | -0.1897 | 0.0595 |
| $8 \times 10^9$ | 0.9774 | 1.160 | 0.2641 |
| $1.6 \times 10^{10}$ | 0.9768 | 1.095 | 0.8332 |
| $3.2 \times 10^{10}$ | 0.9788 | 1.021 | 1.031 |